[A Book Draft]

# Coping with Prospective Memory Failures: An Optimal Reminder System Design

by

Jinghua Hou

Jan 23, 2016

# Abstract

Forgetting is in common in daily life, and 50-80% everyday's forgetting is due to prospective memory failures, which have significant impacts on our life. More seriously, some of these memory lapses can bring fatal consequences such as forgetting a sleeping infant in the back seat of a car. People tend to use various techniques to improve their prospective memory performance. Setting up a reminder is one of the most important techniques. The existing studies provide evidences in support of using reminders to cope with prospective memory failures. However, people are not satisfied with existing reminders because of their limitations in different aspects including reliability, optimization, and adaption.

Through analysing the functions and features of existing reminder systems, this book draft summarizes their advantages and limitations. We are motivated to improve the performance of reminder systems. For the improvements, the relevant theories and mechanisms of prospective memory from psychology must be complied with, incorporated, and applied in this new study. Therefore, prospective memory processes, prospective memory components, and potential factors that influence prospective memory performance were also reviewed in this book draft.

Based on the literature review, a new reminder model is proposed, which includes a novel reminder planer, a prospective memory based agent, and a personalized user model. The reminder planer is responsible for determining the optimal reminder plan (including the optimal number of reminders, the optimal reminding schedule and the optimal reminding way). The prospective memory agent is responsible for executing the reminding processes. The personalized user model is proposed to learn from users' behaviors and preferences based on human-system interactions and is responsible for adapting the reminder plan to meet users' preferences as much as possible. To realize the functions of different components in our new reminder model, a series of principles and algorithms are presented in the book draft.

Our reminder system with the reminder planer, the prospective memory agent, and the personalized user model provides a promising ground for improving the prospective memory performance. The reminder plan, as the product of the reminder planer including the optimal number of reminders, the optimal reminding schedule and the optimal reminding way, overcomes the limitations of current reminder systems in reliability and optimization. Our prospective memory agent supported by psychological models provides a believable diagram of prospective memory processes. It not only demonstrates important components such as planning and monitoring in processes, but also interprets the important stage for prospective memory failures. The execution of reminding integrated into the agent ensures the reminding happens in prospective memory processes to resist prospective memory failures. On the premise of reliability, the personalized user model also acts on the reminder plan to make our system adaptive through learning users' behaviors from human-system interactions.

The future work is required to design and conduct relevant experiments to examine the degree of each factor (the type of ProM task, user's age, the complexity of the ongoing task, the importance of the ProM task, and the motivation of the ProM task ) and the interactions between these factors. We will apply these results into the system to maximize the probability of remembering to perform the prospective task and minimize the potential annoyance. At the same time, we will improve the methods about learning the users' behaviors and preferences from human-system interactions.

# Contents

3 Our ProM Reminder System 25

4 Conclusions and Future Work 42

# List of Figures

# List of Tables

# Chapter 1

# Introduction

## 1.1 Background and Motivation

Forgetting is common in our everyday life. You may forget to do something at a particular time, such as forgetting to take medicines after meal and forgetting a meeting at 3:00 PM, or at a particular occasion, such as forgetting to buy milk when you are passing by a grocery store. These tasks have in common is that they are in the future. Remembering to perform a future task is referred to as prospective memory (ProM) [1]. According to Kliegel and Martin (2003), everyone is undergoing ProM failures. A significant number of 50-80% everyday forgetting is due to ProM failures [2]. Some of these memory lapses can bring fatal consequences. Every year we can see the news that an infant died in a hot car after parents left the car, forgetting that the child was sleeping quietly in the back seat. People are eager to learn more techniques for improving ProM performance and coping with ProM failures.

ProM is a complex cognitive function as it consists of several stages and various components [1, 3]. A specific prospective task has its own characteristics, which determine the nature of the ProM task. First of all, the prospective task to be performed at a specific situation or at a particular time determines the type of ProM (event-based or time-based). Second, the prospect of attending a stressful and exhausting meeting may not be appealing, while the prospect of meeting a friend could be something to look forward to. The motivation is associated with the performance of prospective tasks. However, if the stressful meeting is important since your absence may lead to serious consequences, while the task of meeting a friend may not be so important. The personal awareness of task importance also influences the ProM performance. On the other hand, a ProM task is embedded in ongoing activities, and requires us to perform it when an appropriate condition is satisfied. The degree of attention resources occupied by ongoing tasks could influence the prospective task performance. Improving the ProM performance requires understanding ProM from its mechanisms to various factors that act on it.

ProM problems are common with age [4]. Old adults tend to use ProM reminders to cope with their ProM decline [5]. Without reminders, the challenge for performing a ProM task is that the intention needs to be initiated while you are simultaneously engaging in the ongoing tasks [6]. Reminders provide a solution to initiate the prospective task at an intended time, and they range from paper notes to advanced technology-based ones. Originally, most of technology-based reminders (e.g., MEMOS, Memojog) were designed for users with cognitive impairment to promote their independence and assist them in health and wellbeing of individual [7]. Currently, technology-based reminders are popularly used by individuals. We acknowledge their helps in personal time management, such as Google calendar.

## 1.2 Research Questions

In some cases, we are not satisfied with our reminder system because of its failure in reminding, issuing burdensome reminders (annoyance), or disagreeable signal (e.g., some persons prefer sound reminder to visual reminder). We are motivated to investigate the improvements to the generic reminder system that will be more reliable, optimal and adaptive. Therefore, the tradeoff between the reliability and the annoyance as how many reminders should be issued to users for a specific ProM task is one of our main objectives. Spontaneously, the question of when to remind arises, since reminders are issued between the time of the first reminder and the time of executing the ProM task. Considering the reminder as a created cue to the ProM task, another question of how to make the reminder salient and associated with the ProM task comes up. Therefore, how many times to remind, when to remind, and how to remind are our research questions. Consequently, the number of reminders, the reminding schedule, and the reminding way constitute our reminder plan.

In addition, we learn from some existing intelligent memory assists, such as Autominder with the adaptive feature[7]. We proposed to develop a personalized user model to observe the user's behaviors, learn the user's preferences for each feature of the reminder plan (e.g., the preference of audio or visual signal), and adapt the reminder plan to meet the personal preferences as much as possible. Therefore, our last research question is how to make the reminder plan adaptive and personalized according to the knowledge learned from users.

Based on the discussion in this section above, the highlights of four research questions are:

- How to calculate the optimal number of reminders for a specific ProM task?
- How to determine the optimal reminding schedule for a specific ProM task?
- How to determine the optimal reminding way for a specific ProM task?
- How to make reminder plan adaptive and personalized based on the knowledge learned from users?

## 1.3 Approaches

To address research questions above, our approaches include two dependent aspects. Firstly, we establish a structural model of the reminder system, which includes three components of Reminder Planer, ProM Agent, and Personalized User Model. The reminder planer is responsible for determining the optimal reminder plan. The memory agent is responsible for executing the reminding processes. The personalized users model is responsible for adapting the reminder plan. This structural model is developed by reasoning various factors which potentially influence the ProM performance, by complying with the process model of ProM, and by overcoming the limitations and learning the advantages from existing reminder systems. Secondly, we use mathematical functions to calculate the optimal number of reminders, the reminding schedule and the reminding way. We are also going to apply some machine learning algorithms into the personalized user model to adapt the reminder plan in the future.

## 1.4 Contributions

The main contributions based on literature review, model design and implementation, are summarized as follows:

•	We have done a comprehensive literature review, including three aspects: 1)ProM theories of four stages and six components, 2) a series of potential factors (type, age, the complexity of ongoing task, the importance of ProM task, and the motivation of ProM task) which influence the ProM performance, and 3) the existing reminder systems to cope with ProM failures, all of which provide theatrical and practical supports for our reminder system development.

•	Based on the literature review, we conducted the structural model of reminder system, which includes Reminder Planer, ProM Agent, and Personalized User Model.

•	We came up with the concept of the reminder plan with the optimal number of reminders, the optimal reminding schedule, and the optimal reminding way, which can provide reliable and optimal reminders.

•	We have integrated the reminding function into ProM processes so as to clearly understand how a ProM task fails and how reminders happen in prospective memory processes.

•	We proposed the personalized user model to adapt the reminder plan to meet with users' preferences.

## 1.5 Organization

In the following chapters, firstly, we investigate various factors which potentially influence the ProM performance through reviewing the ProM theories and empirical studies. Secondly, we review a series of existing ProM assists, and compare their advantages and drawbacks. Thirdly, through applying theoretical knowledge, incorporating factors drawn from empirical studies, and learning from the existing ProM assists, we develop our reminder system including three components of the reminder planer, the ProM agent, and the personalized user model. Finally, our future work is proposed to conduct a series of experiments to work out the weight of each factor on the ProM performance, and integrate reinforcement learning and supervised learning to make our reminder system adaptive.

# Chapter 2

# Theoretical and Empirical Background

## 2.1 ProM Theory

Prospective Memory (ProM) refers to remembering to perform the intended task after a delay [1]. The intended task is stored in memory and will be executed in the future [8]. Remembering to take medication at 7:00 PM, remembering to buy milk on the way home after work, or remembering to deliver the mails when passing by a mail box are examples of the ProM task. It happens in our health and social life, and directly influences our life quality. According to Kliegel and Martin (2003), a significant number of 50-80% memory failures are ProM problems [2]. The challenge for ProM tasks is that the intention has to be triggered while you are simultaneously engaging in the ongoing task [6].

Process Model: According to Kliegel et al. (2002), a ProM task involves four stages: 1) intention formation (a future task is planned and encoded), 2) intention retention (the intention is maintained and waits for the perception of the target while engaging in ongoing tasks), 3) intention initiation (the moment at which the execution of intention is initiated), 4) intention execution (the maintained task is performed) [3]. From these stages, we can find that in order to completely accomplish a ProM task, people have to remember not only what content is supposed to do, but also when to perform. If people successfully retrieve what they intend to do from memory, less frequently they fail to remember what the task is.

Therefore, people with ProM problems mostly fail to initiate the intention at the appropriate moment.

Six-Component Model: A number of studies devote to investigating the cognitive processes of ProM in order to explain what components are involved in encoding, retention and retrieval [9, 10, 11]. Dobbs and Reeves (1996) have developed a comprehensive model of six-component which are 1) meta-knowledge, 2)planning, 3)monitoring, 4)recalling the content of the intention, 5)compliance, and 6)awareness of output [9]. In this model, it emphasizes that the retrospective memory (the content of intentions) is involved in ProM processes. Meanwhile, it demonstrates that execution functions, such as planning and monitoring, play important roles in ProM. Additionally, it mentions that both the knowledge of understanding how to remember and personal abilities as potential factors affect the ProM performance.

## 2.2 Factors in ProM

### 2.2.1 The Type of ProM as A Factor

ProM tasks have been primarily identified by two types: time-based and event-based [12]. The time-based ProM refers to performing an intention at a specific time or after a period (e.g., remembering to take medicines at 7:00 PM or 30 minutes later). It depends on internal cues of time monitoring and self-initiation. The event-based task refers to performing the delay intention by an event (e.g., remembering to deliver the mails when passing by a mail box). This event is external environment cue, not requiring self-initiation.

A thorough review of existing studies makes clear that there are mechanical differences between event-based and time-based ProM. A majority of researchers stated that performance on event-based tasks is much better than that on time-based tasks because the latter is particularly dependent on monitoring of time and self-initiation [13, 14, 15]. For example, in studies of Einstein et al. [14], the experiment 3 involved both time-and event-based tasks. The participants were asked to answer a set of general questions (the ongoing task). In the time-based task, the participants were required to press a keyboard key every 5 minutes, and in the event-based task, they were required to press the keyboard key when they met the question of president. The results indicated that the participants' (both 18-21 years old and 61-78 years old adults) performance on the event-based task was higher. However, some researchers presented the different results, for exapmle, d'Ydewalle et al. conducted a face-identification task in the ongoing activity, without answering general questions which were used by Einstein et al. According to their results, the performance in the time-based task was better than that in the event-based task among old adults (55-81 years) (see also [16]).

In the interest of explaining the discrepant results, d'Ydewalle et al. combined experiments both in studies of Einstein et al. and d'Ydewalle et al. (like [17, 14]). They demonstrated that when participants were involved in the simple problems (low cognition), their performance in the time-based task was higher than that in the event-based task, whereas when they were required to resolve complex problems (high cognition), their performance in the time-based task decreased, particularly for old adults (60-86 years) [18]. d'Ydewalle et al. (2001) reproduced the experiment by 2 × 2 × 2 design (old vs. young, event− vs. time− based task, and low vs. high complexity of the ongoing task). This study confirmed that the complexity of ongoing task is an important factor that influences ProM performance. Their result provides a better explanation to the type-related discrepancy [19].

## 2.2.2 Age as A Factor

### 2.2.2.1 Development of ProM

According to the review, only two remarkable studies have examined the development of ProM on very young children (2-5 years old) [20, 21]. Some studies compared children of very young age and early school age (4-7 years old)(e.g., [22]). And some studies concentrated on schoolchildren development with a relatively short age range of 3 or 4 years, such as Passolunghi et al. (1995) compared the ProM performance between 7 years old and 10 years old children, and Nigro et al. (2002) studied children between 7 years old and 11 years old [20]. A few studies concentrated on schoolchildren development with a relatively wider age range of 5 years (e.g., [23, 24]).

From 2 to 5 years old. Two studies examined the development of ProM in children aging from 2 to 5 years old [20, 21]. These studies found that children have the ProM ability at as early as the age of 2. In Somerville et al. (1983), children from 2 to 4 years old were involved in the deliberate ProM activities. 2-, 3-, and 4-year-old children were arranged to 8 different ProM tasks by their mothers over a period of 2 weeks. These tasks varied in high motivation level such as "remind me to buy candy at the store" and low motivation level like "remind me to bring in washing". The time of delay to carry out the task varied in short (a few minutes) and long (a few hours) level. The results found that even 2 years old children could recall the ProM tasks and perform very well, with 80% successful in remembering tasks with high interesting and short delay.

Unlike the naturalistic study of Somerville et al. (1983), Guajardo and Best (2000) studied the 3 to 5-year-old children on the ProM task in a laboratory settings. In their study, the children were introduced a computer-based game in which they received 6 blocks of 10 pictures at frequency of 5 seconds per picture. The ProM task required the children to press a key on the keyboard once they saw a picture of a house (or a duck). The results showed that the 5-year-old children were reliably better at remembering to press the keyboard than 3-year-old children. Their study obtained a significant effect of age which means the 5-year-old children can performance better on ProM tasks than 3-year-old children [25].

From 5 to 7 years old. Some studies showed ProM has rapidly developed after the age of 5 (e.g., [26, 22]). For example, Kvavilashvili et al. (2001) examined two groups of children's (5 years old and 7 year old) performance on the event-based ProM task in experiment 1. The children were engaged in an ongoing activity of naming the picture cards. They were asked to hide the cards when they saw the animal cards. Finally they found that the 7-year-old children performed better than 5-year-old children [22]. However, some other studies obtained discrepant findings. For example, Meacham and Colombo (1980) asked the children (from 5 to 8 years old) to play a card game with experimenters. The ProM task was that when they finished the game, the children reminded the experimenters to open the surprise box which had been placed on the table before they started the game. The data analysis revealed that there was no age effect during this age period [27].

From 7 to 12 years. There are two studies on children development in the time-based ProM from 7 to 12 years old [23, 24]. The computer game CyberCruiser was applied in Kerns' (2000) study. The children from 7 to 12 years old controlled the car by using a joystick. They were required to be very careful about the traffic and hazards around to get points. The points were calculated by whether hitting other vehicles or not and how fast the speed was. The time-based ProM task was to check the fuel gauge by hitting a button in case the gas ran out. The results revealed a significant ProM development from 7 to 12 years old [23]. In the study of Mackinlay et al. (2009), the ongoing activity was one-back picture which required the children (7-12 years old) to judge whether the current picture has been seen before by pressing yes or no keys. The time-based PROM task was instructed to remember to press the clock key every 2 minutes. The results indicated that older children had better performance [24]. Passolunghi et al. (1995) also acquired the findings that the older children's (10-11 years old) performance on the event-based ProM task is better than the younger children's (7-8 years old) [28]. These findings on development patterns revealed a significant ProM development in schoolchildren.

However, Nigro et al. (2002) failed to find any significant development in either time-or event-based ProM [20]. In their study, the children were engaged in the ongoing task of solving mathematical additions and puzzles. They were required to remind the experimenter to do something at a particular time or when seeing another experimenter. The results didn't present age effect between 7-year-old and 11-year-old children. Kvavilashivili et al. (2007) concluded that development of ProM is slow from 7 to 12 years old [29]. This conclusion contradicted the findings in Mackinlay et al. (2009) which showed significant ProM development in the time-based task, and the findings of Passolunghi et al. (1995) which showed the children's ProM development in the event-based task.

Conclusions: Why the findings of these studies are discrepant? Firstly, as mentioned before, the complexity of the ongoing task possibly interfered with the results. For examples, in the study of Guajardo and Best (2000), each block of 10 pictures with 5 seconds per picture to response could have been a difficult ongoing task for 3-year-old children. Also, in the study of Kvavilashvili et al. (2001), it is obvious that 4-year-old children have more challenges to name 20 pictures in each of four stacks of cards than children of 7-year-old. Similarly, in the study of Kerns (2000), although the game as an ongoing task was equally interesting to both 7 and 12-year-old children, it is still difficult for 7-year-old children to play that game.

Secondly, it is also possible that some paradigms cannot eliminate ceiling effects. For examples, in the study of Somerville et al. (1983), the ProM task of reminding mom to buy candies in a few minutes was too exciting for either 2-year-old children or 3-year-old children. The children cannot wait to carry out the exciting task (ProM task). It is not surprising to see 2-year-old children's performance on the ProM task is the same as older children's. Similarly, in the study of Meacham and Colombo (1980), the ProM task of reminding the experimenter to open the surprising box is also a temptation target.

2.2.2.2 Aging with ProM

In fact, initially, age was the focus of ProM studies. Craik (1986) was the first one to point out that age would be a factor that effects on ProM performance. In his study, they found that the older adults have the more complaints about their forgetfulness in terms of accomplishing deferred tasks [13]. Later, Dobbs and Rule (1987) examined age differences in ProM performance [30]. They also found a deficit for older adults (from 70 to 99 years old) compared with younger adults (from 30 to 65 years old).

Taking ProM types into account, Einstein, et al. (1995) pointed out that age effect exists on the time-based ProM by comparing old-adults and young-adults. At the same time, they stated that there was no age effect on event-based ProM [14]. d'Ydewalle et al. argued against their results through a set of experiments [18]. d'Ydewalle (1996) also conducted three experiments to test the performance on the time-and the event-based ProM task [16]. Unlike Einstein et al. (1995), they found that there was age effects existed in the event-based task.

To address the discrepancy of the event-based ProM, d'Ydewalle et al. (1999) applied two types of ongoing activities into their experiments which included answering general knowledge questions used by Einstein et al. (1995) and face-identification used by dYdewalle et al. (1996) [18]. The answering-questions activity is more complex than face-identification because the former requires more execution functions. The result showed that old adults' performance in both the time-and event-based task declined regardless of ongoing activities. d'Ydewalle et al. confirmed that age effect exists in both the time-and event-based ProM.

Additionally, Maylor (1996) also conducted experiments to examine the discrepancies. She pointed out the difficulty of ongoing activities might be an interference factor to detect the age-related difference [4]. In other words, if the difficulty of the ongoing task is manipulated in the experiment, it would equate ProM performance between younger and older participants, such as Einstein and McDaniel's (1990) study, in which they controlled the older participants' ongoing task with fewer words in the list of recalled words. Maylor's (1996) experiment asked participants to recognize the face and give response when they found the person had a beard. This experiment required the attention shifting from face-recognition (one level of processing) to features of face (another level of processing). The attention shifting is a process of execution functions. Compared with younger adults, older adults have diminished execution functions [31]. Therefore, Maylor (1996) concluded that older adults also have deficits in the event-based ProM task which requires execution functions.

Look through the ProM development across the lifespan, compared with young adults, children have less developed ProM and older adults have gradually declined in ProM, which is attributed to the development of execution functions. These age differences draw an inverted U-shaped developmental trajectory across lifespan as replying on measures of the execution function [1].

Conclusions: It is clear that age-related differences in the time-based ProM performance is related to internal cues of time monitoring and self-initiated. Age-related differences in the event-based ProM depend on some variables of execution functions such as the attention shifting from different levels of cognitive processing [4]. Therefore, both the time-based and event-based ProM seem to require the executive control processes [3, 32, 33].

In the laboratory setting, the ongoing task is manipulated by experimenters. Failure of adjusting the difficulty of ongoing task across the age groups may lead to confusing and contradictory results such as what we mentioned above.

### 2.2.3 The Complexity of the Ongoing Task as A Factor

In laboratory settings, the ongoing task is designed as a background work such as answering a set of questions [14], solving the mathematical or puzzle problems [20], naming a set of pictures [16, 18, 22, 4], or playing a computer-based games [23], etc., from which the participants need to perform ProM tasks by shifting their attention from the ongoing task.

Firstly, take a look at some studies of comparing ProM performances between younger adults and older adults. In studies of Einstein et al. (1995), the ongoing task requires participants to answer a long set of general knowledge and problems-solving questions both in the time-based and event-based ProM [14]. d'Ydewalle et al. (1996, 1999) employed the paradigm similar to those in Einstein et al. (1995), but face-identification as a different ongoing task applied in the paradigm [16, 18]. Similarly, in the study of Maylor (1996), the ongoing task was designed as writing down the name of various famous male faces [4].

In the developmental studies of ProM, Nigro et al. (2002) designed their ongoing task as solving a series of mathematical operations and puzzles for children. The difficulty of the ongoing task was previously adjusted according to the age of the participants both in the time-based and event-based ProM [20]. In the study of Kvavilashvili et al. (2001), the children aged 4, 5, 7 years were asked to look at four stacks of cards with picture one by one and tell the experimenter as accurately as possible what the picture is. The children were also told that they can draw one picture for each stack [22]. Kern (2001) employed computer-based game in the experiment. The ongoing task required the children aged 7 to 12 years to play the Cyber Cruiser game, in which the children used a joystick to control a car in a road with traffic and hazards [23].

The naturalistic studies carry out experiments in the real-life environment. The study of Somerville et al. (1983) is a pioneer of naturalistic study. In this study, children aged from 2 to 4 years were given eight ProM tasks in their real life by their mothers. Their mothers also observed their ProM performance and acted as an experimenter [21]. Ceci and Bronfennbrenner (1985) applied a video game to mimic naturalistic study with a real-life scenario. The children were allowed to play the video game, but they needed to remember to take the cupcakes out of the oven in a delayed time of 30 minutes [34]. The ongoing task of playing the game is very likely to happen in our life. It is the same as watching TV but remembering to turn off the oven after certain minutes. In the study of Kvavilashvili and Fisher (2007), the participants did what they did daily as usual. They were required to remember to make phone calls to the experimenter either at a pre-arranged time or after receiving a certain text message on the seventh days of their experiment session [29]. The ongoing task was also their real life, in which the ProM task was embedded as making a phone call at an intended time.

The performance of the ongoing task. Most of ProM studies only pay attention on ProM performance, whereas the performance on the ongoing task is not reported. For instance, Kern (2000) emphasized that the computer-based game of Cyber Cruiser made children aged from 7 to 13 years equally engaged in playing, but Kern didn't analyse the performance of ongoing task (game score) to identify the age-related difference. Although Nigro et al. (2002) previously adjusted the complexity of the ongoing task (mathematical problems and puzzles) according to the class level for the children aged from 7 to 12 years, they didn't report children's performance on the ongoing task.

Few studies reported the performance of ongoing task [18, 35, 36]. d'Ydewalle et al. (1999), for example, they conducted the experiment by 2 × 2 × 2 design (old vs. young, event− vs. time− based task, and low vs. high complexity of the ongoing task). In their results, they analysed the ongoing task of both answering questions and face identification. Under the condition of answering questions (high complexity), older adults even performed better on the ongoing task than younger adults in the time-based task, whereas there is no age difference in the event-based task. On the other hand, under the condition of face identification (low complexity), younger adults performed better on the ongoing task than older adults both in the time-based and event-based task, but the age-related difference in the event-based task is not significant. The analysis of interactions (ProM and ongoing task; ProM and age; age and ongoing task) suggested that age differences in ProM would disappear under some conditions (low complexity of the ongoing task) when taking the performance of ongoing task into account [18].

The study of Rendell et al. (2007) involved the face recognition as the ongoing task, in which a target cue "with glasses" or "John" is embedded. The ongoing task was measured by the proportion of famous faces correctly named. In experiment 2, the ongoing task was slowly paced than in experiment 1. They found that younger participants named a greater proportion of faces than the older adults did in both experiment 1 and experiment 2. More importantly, the finding patterns between experiment 1 and 2 indicated that when the ongoing task is less challenging (experiment 2), the older adults performed on ProM as well as the younger adults [35].

Smith et al. (2013) embedded a ProM task in an ongoing color-matching task. They manipulated the difficulty of the ongoing task by varying the number of colors, such that older adults performed a 4-color version of the color-matching task, while younger adults did the 4-color version of color-matching task in one trial and a 6-color one in another trial. They found that even if older adults, as well as younger adults, failed to perform the ProM task, they performed similarly or better than younger adults on the ongoing task. They also stated that emphasizing importance of the ongoing task can improve the ongoing task performance [36].

Conclusions: On the one hand, the studies have already paid attention to the complexity of the ongoing task. On the other hand, the studies indicated that the complexity of the ongoing task is an important factor to interfere age-related, and type-related differences.

### 2.2.4 The Importance of the ProM Task as A Factor

The existing research supports that the importance of ProM task can affect the ProM performance, such that ProM performance would improve when the ProM task makes individual to allocate increased attentions [37, 36]. Kliegal et al. (2001), in order to understand the importance of ProM task, they conducted two experiments, in which the task importance was manipulated. They found that the ProM performance improved with the higher importance of the ProM task.

Smith et al. (2013) study also investigated the effects of the task importance. In this study, one group of participants received instructions emphasizing the importance of the ProM task (PMI) compared with another group of participants who received instructions emphasizing the importance of the ongoing color-matching task (CMI). The results showed that participants performed better on the ProM task under the PMI coniditon. More interesting, the age-related differences exist in the PMI condition: for younger adults, emphasizing the importance of ProM task substantially improve their performance on the ProM task and decrease their performance on the ongoing task, whereas older adults in the PMI condition slightly improve their ProM performance and kept the same level of performance on the ongoing task [36]. In other words, younger adults can vary their allocation of resources between the ongoing task and the ProM task as a function of task emphasis, whereas old adults are less capable of shifting their attention from the ongoing task to the ProM task, and they assumed that the ongoing task is more important for them than the ProM task.

Conclusions: Nonetheless, we clearly understand that emphasizing the importance of ProM task improves the ProM performance, although the degree of improvement varies by age.

### 2.2.5 The Motivation of the ProM Task as A Factor

Some studies have examined the effect of motivation to the ProM performance, especially in the naturalistic study [21].

The study of Meacham and Singer (1977) was the first one to investigate the motivation to ProM. In their study, the participants were instructed to send postcards to the experimenter on the rearranged time. They results revealed that the participants who expected to receive a reward performed better than those who did not expect to receive a reward [38].

In the study of Somerville et al., (1983), for an example, the children from 2 to 4 years old involved in the deliberate ProM activities. They were given 8 different ProM tasks by their mothers over a period of 2 weeks. These ProM tasks varied in high motivation level such as "remind me to buy candy at the store" and low motivation level like "remind me to bring in washing". The results found that even 2-year-old children could recall the ProM tasks with 80% successful in remembering high motivation tasks, as well as 4-year-old children [21].

Recently, in order to identify the motivation as an assumed factor for the paradox of age-related declines in laboratory compared with age benefits in naturalistic settings, Aberle et al. (2010) conducted an experiment, in which participants were instructed to remember to contact the experimenter repeatedly over the course of one week. One group has monetary incentive and the other group has no incentive. The results showed that young adults in the high motivation group overcame their age-related deficits and performed better than those in the low motivation group [39]. In other words, increasing the motivation of ProM task can improve the ProM performance.

## 2.3 ProM Assists

As mentioned before, ProM is vital for our health and social life. ProM failures produce great challenges to people and directly influence their life quality. According to Kliegel and Martin (2003), a significant number of 50-80% memory failures are ProM problems [2]. To avoid the consequences of ProM failures, people are likely to use memory assists to help their remembering, especially for old people [40]. Most studies demonstrated that both young and old people benefit from using memory assists (e.g., [12, 41]). Memory assists help users to store information or to remind the user an event they might forget [42]. In this study, we focus on the reminder function of memory assists relative to ProM, rather than the storing function relative to retrospective memory.

According to Harris (1978), reminders are generally categorized as active or passive reminders [43]. Examples of diaries, lists, and calendars are passive reminders which require the user to actively check them, whereas google calendar and mobile phone are examples of active reminders which attract users' attention and instruct them when and how to perform an intention. ProM reminders vary from traditional way of pen and paper to technology-based way of electronic devices. The purpose of designing a ProM reminder also varies from specific to generic use. The current study targets on a technology-based reminder of generic use, since our ultimate aim is to produce a reminder system with more flexible and adaptable features.

### 2.3.1 Current ProM Assists

#### 2.3.1.1 NeuroPage

Originally, the technology-based memory assists have commonly targeted on cognitively impaired people. One of the earliest ProM assist was Neuropage, of which the primary users are brain injured patients [44]. Neuropage is very simple with little learning. Users enter the schedule information through a paging company, and reminder alerts are sent to the user at an appropriate time. These functions ensure NeuroPage is a high usable ProM assist. Wilson et al. (1997) evaluated the NeuroPage. The result showed that NeuroPage as a ProM assist improved participants (19-66 years old) performance on ProM tasks such as remembering to take medicines or pack lunch [45].

However, according to a study of Caprani et al. (2006), there are two potential functions relating to ProM NeuroPage can improve. The first one is the postpone function. The second one is the task confirmation function [7]. A user may receive a reminder at an unsuitable time, therefore the ProM assist has a better function of task postponement so that the user can be reminded at a suitable time when they are available to successfully perform the intention. Meanwhile, the caregiver should know whether the intended task has been carried out or not. Involving these two functions can improve NeuroPage ability of the assistance.

2.3.1.2 MEMOS (Mobile Extensible Memory Aid System)

MEMOS is a mobile interactive ProM assist, which was designed for brain injured and older users to remind them of essential facts and dates [46]. MEMOS consists two parts of personal memory assistant (PMA) and a base station. PMA is a mobile electronic device to remind the user of important tasks and provide feedback, and the base station coordinates the activities of caregivers and notifies them about feedback of the task execution.

Walther et al. (2004) evaluated the MEMOS. The participants were patients with head injury. The result showed that users performed highest in using PMA from MEMOS compared with other electronic memory assists (palm pilot and mobile phone) [47]. MEMOS overcomes the major drawbacks of the NeuroPage. It supports caregivers to encode and input information, and displays important information for patients to successfully perform the task. Furthermore, it allows patients to confirm the task carried out by pressing a button, and PMA can detect this information.

Similar to NeuroPage, the primary users of MEMOS are brain injured patients. The research group has recognized its potential ProM assists for healthy older people, and planned to extend the system for an application in other field [48].

2.3.1.3 Memojog

Memojog was designed as a ProM assist built in a personal digital assistant (PDA) platform for memory impaired persons [49]. Memojog consists three components of PDA, the central sever, and the web-based database. User, caregiver, or care professional can input the users schedule and action prompts in the PDA or web-based database. The user can accept, postpone or ignore the reminder. The caregiver and care professional can acquire the user's response by data transmitted to the central server. Memojog also has multiple functions of storing personal information for the user.

Memojog system was evaluated with a group of old adults and memory-impaired users [50]. There were two field evaluations comprising of 6 participants in each evaluation (different participants in each evaluation). The results showed that the participants were happy with the Memojog system and could use it easily. Users appreciated the system reminded their intended tasks accurately.

However, the participant also gave some negative comments such as coverage problem of the inability to connect to the relevant website for changing or updating their schedule, and some hardware-related problems, e.g., the touch screen was not sensitive [50].

2.3.1.4 Google Calendar

Both paper-based and electronic-based calendar are used. Paper-based calendars have a few limitations associated with their use, such as no alerts, forgetting to look at it and no enough space. Electronic-based calendar, such as Google Calendar, is an alternative solution to overcome these limitations. Google Calendar not only provides email reminders and pop-up reminders, but also enables users to link their calendar with their mobile phones. Users can set how far in advance of the task and how many reminders, which enable users to setup the system according to their personal needs [51]. Google calendar is also a simple system to use, such as clicking on the box corresponding to the appropriate date and time to create an event, filling the key words in the blank to set up what the event is.

McDonald et al. (2011) evaluated the effectiveness of Goolge Calendar by comparing the standard diary use. The results showed that the participants (19-65 years old) completed more intentions when their using Google Calendar than when their using standard diaries. In other words, Google calendar is more effective than the diary to support people to achieve their ProM tasks. They also found that Google Calendar was rated higher than the standard diary in assist preference. Compared to the diary, Google Calendar reduces the need for monitoring by alerting the participants to complete the events; the active reminder (linking to mobile phone) reduces the need for actively or frequently checking the calendar. Therefore, Google Calendar appeared to support the retrieval of ProM tasks, and maximize the probability that ProM tasks can be carried out within the necessary response window [51].

However, the participants in McDanald et al. (2006) complained that they finally failed in the task execution even if they noticed a timed reminder to perform a task. The main reason is that the reminder issued at users' unavailable time, but reminders cannot interact with users to update the reminders in real-time.

### 2.3.1.5 AutoMinder

Autominder is one of the most advanced technological reminder systems, which assists a broad population such as older adults in general [52]. The purpose of Autominder is to help the older adults live perfectly in their home environment. Compared to the previous ProM assists, Autominder has the ability to model user's daily plans, track user's task execution through the behaviors detected by the sensors at home, and make decisions about whether and when to issue reminders [52].

Autominder has three main components [52]. The first component is Plan Manager, which is responsible for storing user's initialized daily plan and updating the user's plan as the day progresses to avoid the inconsistent/conflicting activities. The second component is Client Modeler, which is responsible for monitoring the execution of the user's plan through the information of observable behaviors, as well as knowledge of whether and when any reminders were issued. The third component is Personalized Cognitive Orthotic, which is responsible for deciding what and when to issue the reminder. These three components illustrate Autominder's main functions. Its impressive abilities provide older adults a intelligent and smart living home.

However, according to Caprani et al. (2006), the sensors and observable information are not always reliable, which may result in assumption failures. Assumption failures decrease the reliability of reminders. Furthermore, the old users may be afraid of intelligent technology, for example, they may feel wary of a mobile robot and sensors working at home [7].

### 2.3.2 Comparisons of ProM Assists

| Assist | NeuroPage | MEMOS | Memojog | Google Calendar | Autominder |
|---|---|---|---|---|---|
| Function | Alerts users to perform tasks | Alert users to perform future tasks | Alert users to perform future tasks. Hold personal informa | Alert users to perform future tasks. Hold events | Adaptive reminder that alters schedule based on users past actions |

Based on a review paper [7], we list functions and memory supports of several electronic ProM assists.

Each memory assist has their evaluation studies to support their effective use of helping users to execute prospective tasks. From the review above, they have their own advantages and limitations. We compare their relevance on each stage of ProM theory: 1) intention formation: All of them support encoding, as they require users to form intentions. Some of them provide voice inputs (e.g., MEMOS, Autominder). 2) intention retention: All of them support short or long term delays, unlike the retention time in human memory which is limited. However, most of them are restricted to the time-based ProM, and only Autominder supports both time-based and event-based ProM. 3) intention initiation: All of them can trigger user's intentions to reduce the need for monitoring. 4) intention execution: Their text-based prompts help users to execute ProM tasks.

Table 2.1: Comparisons of ProM Assists

# Chapter 3 Our ProM Reminder System

## 3.1 Introduction

Based on the literature review in chapter 2, we find that 1) support all stages of the process model [3], and 2) meet the requirement of ProM components (e.g., planning, monitoring, content recall). However, part of prospective tasks is still not carried out successfully are mainly due to the reminder issued at users' unavailable time and the reminder system failing in reminding.

Some ProM assists can provide multiple reminders, such as Google Calendar and users can increase the number of reminders by manually changing the setting. Increasing the number of reminders is a solution to increase probability of reminders issued at a user's available time. In this case, how many reminders should be issued to a user? One time reminder is not enough to guarantee the completion of a ProM task. Too many reminders may lead to annoyance.

We are motivated to investigate how many reminders should be present for a specific ProM task which has not been addressed by existing approaches. Spontaneously, the question of when to remind arises because reminders are issued between the time of starting the reminder and the time of executing the ProM task. Being back to the reminder, the nature of it is a deliberated cue for a ProM task. So the third question is how to make this cue (reminder) salient, and highly associated with the ProM task.

Therefore, we propose to develop a reminder plan to solve these three questions. In the following we explain the three outcomes of this reminder plan including the optimal number of reminders, the optimal reminding schedule, and the optimal reminding way.

1 The optimal number of reminders depends on factors which have effects on the ProM performance. Although Autominder, Client has an optimization function which reason the user's behavior patterns to decide whether and when to remind, the observable information is not always reliable. Particularly for old people, advanced and intelligent technology has already made them to feel uncomfortable and unsafe. Even worse, assumption failures could cause confusion and apprehension [7]. Therefore, the optimal number of reminders determined by the inputs of factors should avoid system assumption failures and annoyances of redundant reminders.

2 The optimal reminding schedule decides when to issue reminders before the ProM task. The first factor is users' assumption of how long they will be ready for the intended task. For example, you have a meeting at 2:00 PM in auditorium. You are planning to work at office before the meeting. If you assume it will take you 30 minutes to get to auditorium, it is more likely that you setup the reminder at 1:20 PM. If you assume it will take you only 5 minutes to get there, it is more likely that you setup the reminder at 1:50 PM. The second factor is retention intervals of the ProM task, because intervals range from hours to months. Even if there are multiple reminders for a ProM task which needs to be executed 7 days later, you don't want to receive reminders from today. The third factor is users' current location and situation. For example, if you setup the reminder at 1:30 PM for the

2:00 PM meeting, actually, the time cost on distance between the meeting place and your current place is much longer than 30 minutes. In this case, we proposed to adjust the reminding schedule earlier according to the actual location-based time cost.

26 3 The optimal reminding way. The nature of reminder is a deliberately created cue. Several studies support that cues strongly associated with the intention can produce effectively retrieval [53, 54]. Reminders with simple alarms/alerts cannot support the cue strongly associate with a specific ProM task. Users may encounter confusion if all ProM tasks are reminded by the same sound, especially when visual reminders are unavailable or unreachable. Our reminder system can understand the type of the ProM task (e.g., doctor appointment) to produce high associated reminders.

At the same time, because of the interplay between the ProM task and the ongoing task, we also propose that our reminder system should understand the setting of the ongoing task. For example, when you are engaging in an important meeting and you have a doctor appointment 30 minutes later, normally you don't want to be reminded loudly which is interruptive. In office-based environment, users are more likely to use text-based reminders, such as pop-up windows.

However, how to remind is also an intricate question which is concerned with not only the nature of ProM task and the environmental context, but also individual differences. Some users prefer the audio reminder so as to make sure it is heard. Some users with cognitive impairment may hope the reminder with a picture. The thought of how to remind is consistent with the appeal of reminder systems requiring human factor analysis [7].

After generating the reminder plan, we propose a ProM agent to implement the plan. First, in our ProM agent, the reminder as a target cue appears at the initiation stage. Meanwhile, the reminder plan is also encoded and maintained in storage with the intended task together, waiting for the moment to be issued. Second, the ProM agent issues the reminders, and determines the next step based on the user's response to the reminder.

Finally, the ProM agent supports updates before and after the reminding process.

Finally, ProM assists should be adaptive. Through observing the user's behaviors, the system can reason and determine whether and when to issue a reminder to users. We propose to develop a personalized user model to observe the user's behaviors, learn the user's preferences for each feature of the reminder plan and adapt the reminder plan to meet the personal preferences as much as possible.

The current study targets on a technology-based reminder of generic use, since our ultimate objective is to produce a reminder system with more flexible and adaptable features.

## 3.2 Key Ideas for Developing Our Reminder System

The research problems we discussed above need to be addressed on a level of theory, model and practical knowledge based on analysis of the relevant psychological theories and limitations of currently used reminder systems. To design our reminder system, firstly, we overcome the limitations of the existing ProM assists and learn from their advantages. Secondly, we develop a computational reminder model based on the ProM theories. Thirdly, we incorporate and reason the factors which potentially influence the ProM performance.

### 3.2.1 Ideas from Existing ProM Assists

Google Calendar is one of the most popular memory assists. It is welcomed by large population with its easy learn and simple use. However, according to McDonald et al. (2011), participants still failed in some prospective activities in their everyday life even if they used the Google Calendar [51]. As we discussed above, Google calendar as a memory agent supports memory processes of encoding, retention, and retrieval. However, in daily life, there are various situations when receiving a reminder at an inappropriate time, such as the busy situation of meeting. The users cannot carry out the intended task at that time. They hope they can do it later. Therefore a reminder would benefit from a function of postponement so that the users can be reminded at a time when they are available. At the same time, when current situation such as location requires more time to be ready for the ProM task, the reminder would be adjusted earlier than before. Unfortunately, Google Calendar does not provide the function of adjustment the reminders synchronously according to the current situation. The advanced reminder system -Autominder, although it has adaptive feature, it lacks the stage of users' encoding. The observable information is not always reliable, and system assumption failures could cause confusion and apprehension to users [7].

Our reminder system has the features of reliability, optimality and adaptivity to meet each individual's requirements as much as possible.

### 3.2.2 A Computational Model Developed from ProM Theories

With research of ProM gradually growing, several theories have been developed to explain the processes and components involved in ProM. The process model [3] claimed that there are four stages in ProM: 1) formation, 2) retention, 3) initiation, and 4) execution. At the same time, Dobbs and Reeves (1996) have developed a model of six components: 1) metaknowledge, 2) planning, 3) monitoring, 4) content recall, 5) compliance, and 6) awareness of output. These models explain what ProM processes are and how cognitive demanding is. Altering the nature of the ProM task can consequently alter the component necessary to implement the task. For instance, setting an active reminder accompanying with an alert could eliminate the necessary stage of monitoring [55]. The reminder system can reduce the ProM load by helping users to remember intentions (It acts as a memory agent). Therefore, it is important to understand the nature of ProM for developing an effective and reliable reminder system.

    We propose that our reminder system could help users to plan and organize the information clearly. In the encoding (the same as formation) stage, our reminder system requires user input information to avoid the unreliability. However, the system provides the salient and distinctive categories to fill in (e.g., what, when, where, who) so that users can consequently receive cues that highly related to the prospective task. Uniquely, this system encodes the optimal number of reminders, the optimal reminding schedule, and the optimal reminding way. In the retention stage, the system can update the information according to user responses. For examples, the system will set the number of reminders null when users accept the task; the system will maintain the number of reminders and issue reminders later when users postpone the task. The initiation stage is the most important one since people with ProM problems mainly fail to initiate the intention at the appropriate moment [6]. Especially, maintaining the monitoring is more demanding of cognitive resources. A reminder system initiates the task and issues a reminder to users at an appropriate time, which perfectly substitutes human's time-monitoring or cue-capturing. Definitely, our system is capable of eliminating the necessity of the human initiation stage. In the execution stage, if the system detects a response of accepting from terminal, it assumes that the user is carrying out the task. Ideally, users who have cognitive impairment should receive the step-by-step guidelines to ensure that the task will be successfully executed. Our system considers this function for users with cognitive impairment.

### 3.2.3 Incorporate and Reason Factors Affecting the ProM Performance

From the literature review, whether in laboratory or naturalistic settings, we know that various factors that influence the ProM performance, such as the type (time-based vs. event-based), the user's age, the complexity of the ongoing task, the importance and motivation of the ProM task. Some factors are negatively relative to the ProM performance, for example, participants' performance is better when they are involved in simple problems than they are involved in complex problems [17, 18]. The others, like the importance of the ProM task, are positively related to the ProM performance. A lot of studies demonstrated that participants performed better on the ProM task if the importance of the ProM task emphasized [37, 36]. So far, we have no computational model to incorporate all these factors into a memory agent.

In daily life, the situation and the context of ProM is variable and flexible.generally, No factor can guarantee the success of prospective tasks. However, we can maximize the probability of the ProM completion as much as possible. Meanwhile, we can set the optimal number of reminders to obliterate redundance. Consequently, we can achieve a reliable and optimal reminder system. Therefore, we firstly need to incorporate relevant factors into the reminder system, and then work out the weight of each factor. Finally, we use relevant principles to calculate the optimal number of reminders, the optimal reminding schedule, and the optimal reminding way.

## 3.3 Implementation of Our Reminder System

Our reminder system includes three components: the reminder planer, the ProM agent, and the personalized user model (see Figure 3.1). The reminder planer is responsible for producing the reminder plan according to potential factors on ProM. The ProM agent is responsible for encoding the task and the plan, maintaining the task and the plan, initiating the task, and performing the task, where reminders are triggered and issued at the initiation stage. The personalized user model is responsible for adapting the reminder plan according to human-system interactions. In the following sections we discuss the detailed implementation of each component.

### 3.3.1 Modeling of the Reminder Planer

The reminder planer is designed for producing the reminder plan according to potential factors on ProM. The optimal number of reminders, the optimal reminding schedule, and the optimal reminding way are produced from different functions. In the following, we show knowledge representations and the implementation of each function.

Figure 3.1: The Reminder System Model

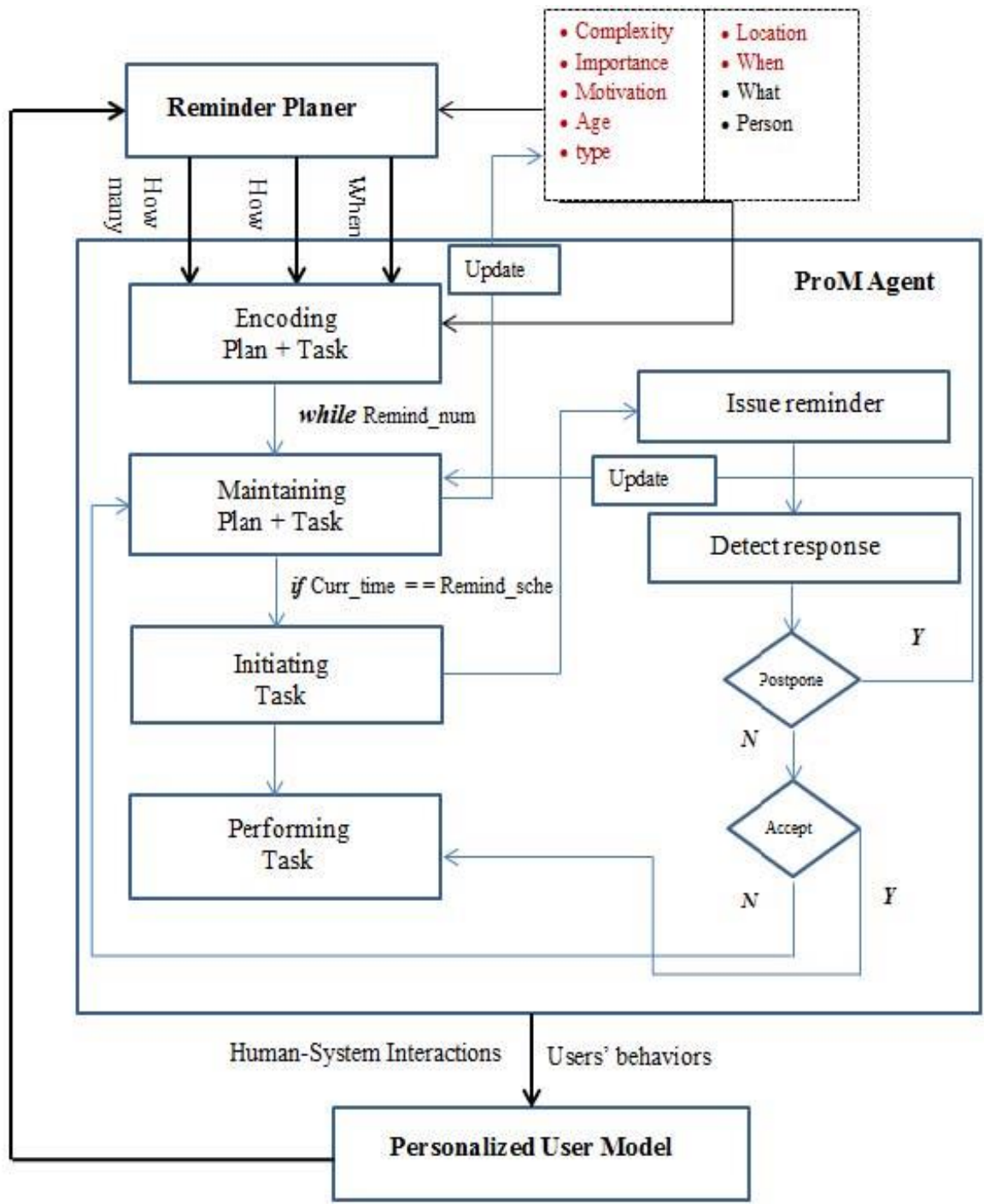

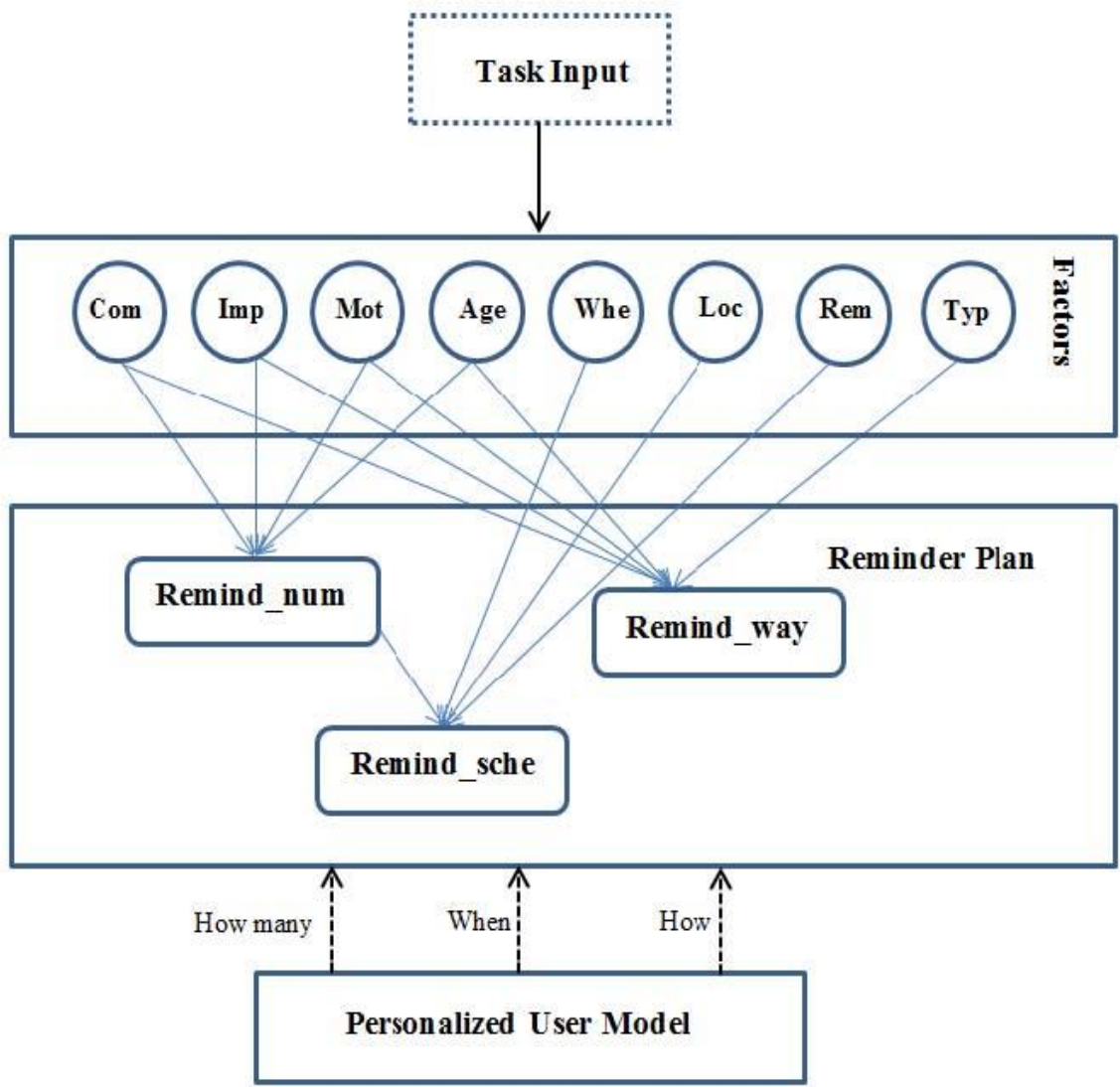

Figure 3.2: The Reminder Planer

The reminder planer has three functions. The first function is to compute the optimal number of reminders. The second function is to compute the optimal reminding schedule.

The third function is to compute the optimal reminding way (see Figure 3.2).

Function -The optimal number of reminders: The Pn(Com, Imp, Mot, Age) is used to compute the optimal number of reminders. This function includes four variables: the complexity of the ongoing task, the importance of the ProM task, the motivation of the ProM task, and the user's age. The possible values for the first three variables are low, medium, and high. The value of the last variable can be either young or old. Each variable has an associated weight to influence the ProM performance. Therefore, we apply the weighted average method to calculate the optimal number of reminders.

Now we formally define the variables and the implementation of this function. The variables are:

- Remind num is the optimal number of reminders;

- Com is the complexity of the ongoing task;

- Imp is the importance of the ProM task;

- Mot is the motivation of the ProM task;

- Age is the user's age;

- The value of Com, Imp, and Mot, is Low, Medium, or High correspondingly denoted by L, M, or H;

- The value of Age is young or old, correspondingly denoted by y and o;

- $W = \{w_1, w_2, w_3, w_4\}$ // W is a set of the weights corresponding to each variable in $P_n$;

- $T = \{t_1, t_2, t_3, t_4\}$ // T is a set of the numbers of reminders corresponding to each variable in $P_n$;

| Variable | Com | Imp | Mot | Age |
|---|---|---|---|---|
| Input | LMH | LMH | LMH | LMH |
| W | $w_1$ | $w_2$ | $w_3$ | $w_4$ |
| T | $t_1$ | $t_2$ | $t_3$ | $t_4$ |

Table 3.1: The number of reminders corresponding to each variable in Pn

Now we define the implementation of the function Pn(Com, Imp, Mot, Age).

Pn (Com, Imp, Mot, Age)

{ constant $n_L, n_M, n_H, a_y, a_o$;

  Principles:

  { if (Com == L), then $t_1 = n_H$ ;

   if (Com == M), then $t_1 = n_M$ ;

   if (Com == H), then $t_1 = n_L$;

   The same principles to Imp, and Mot;

   if (Age == y), then $t_4 = a_y$;

   if (Age == o), then $t_4 = a_o$;

  }

  Remind num = $\frac{\sum_{i=1}^{4} w_i \cdot t_i}{\sum_{i=1}^{4} w_i}$ ;

  Return Remind num;

}

Function -The optimal reminding schedule:

Pt(Loc, Rem, W he, Remind num), this function is to produce the optimal reminding schedule, which consists of each reminding time. The optimal reminding schedule is distributed from the time of the first reminder to the time of executing the ProM task.

The reminder's starting time is mainly determined by user's expectation. But if the location-based time cost is more than interval between the time of the ProM task and the user's expectation of the reminder's starting time, we adjust the reminding schedule earlier by adding the time cost into the interval.

Now we formally define the variables and the implementation of this function. The variables are:

- Loc is the place of the ProM task;
- Rem is the user's input of the first reminder's time;
- W he is the execution time of the ProM task;
- $R = \{r_1, r_2, r_3, \cdots, r_n\}$; // R is a set of reminding time
- Curr loc is defined as the user's current location;
- Dis is defined as the distance between Curr loc and Loc;
- T ime cost is defined as the time need to reach Dis;

Now we define the implementation of the function Pt(Loc, Rem, W he, reminder num).
Pt (Loc, Rem, W he, Remind num) { R distributed on [Rem, W he];
 The number of r is Remind num; Interval = W he -Rem; Principles: { if (T ime cost more than Interval), then
   Rem = W he -(T ime cost + Rem);
  }
  interval = W he -Rem; update R;
}

| | Assist | NeuroPage | MEMOS | Memojog | Google Calendar | Aut r |
|---|---|---|---|---|---|---|
| Function | Alert user | Alert users | Alert users | Alert users | Alert users | Ada |

Please note that Time cost is calculated according to the travel speed (walk: 5km/hour; car: 60km/hour). For example, if Dis is 2.5 km and by walking, then Time cost is 30 minutes.

Function -The optimal reminding way:

Ph(Com, Imp, Mot, Age, Typ), this function is similar to the function Pn for computing. The optimal reminding way is computed by using the weighted average method. Besides the four variables in determining the optimal number of reminders, it also includes another variable -the type of the ProM task to determine the reminding way. The input of each variable produces the corresponding way of reminding. Differently, the input of the task type is classified by personal, financial, social, or work, which also correspondingly links to the reminding way.

Now we formally define the variables and the implementation of this function. The variables are:

- Remind way is the optimal reminding way;
- Typ is the type of the ProM task;
- The value of Com, Imp, Mot is defined as Low, Medium, or High, correspondingly denoted by L, M, or H;
- The value of Age is defined as young or old, correspondingly denoted by y and o;
- The value of Typ is defined as Personal, Finance, Social, and Work, correspondingly denoted by per, fin, soc, and wor;
- How = $\{h_1, h_2, h_3, h_4, h_5\}$ // How is a set of reminding ways (visual vs. audio, long vs. short, or music vs. ring) corresponding to each variable in Ph;
- W = $\{w_1, w_2, w_3, w_4, w_5\}$ // W is a set of weights corresponding to each variable in Ph; 37

Table 3.2: The reminding way corresponding to each variable in Ph

The function Ph (Com, Imp, Mot, Age, Typ) can be defined using the following pseudocode.

{ constant $t_p, t_f, t_s, t_w, a_y, a_o, h_L, h_M, h_H$ ; Principles: { if (Com == L), then $h_1$ is $h_H$ ;

if (Com == M), then $h_1$ is $h_M$ ; if (Com == H), then $h_1$ is $h_L$; The same principles to Imp, and Mot; if (Age == y), then $h_4$ is $a_y$; if (Age == o), then $h_4$ is $a_o$; if (Typ == per), then $h_5$ is $t_p$; if (Typ == fin), then $h_5$ is $t_f$ ; if (Typ == s), then $h_5$ is $t_s$; if (Tye == w), then $h_5$ is $t_w$;

}

Remind way = $\sum_{i=1}^{5} w_i * h_i / \sum_{i=1}^{5} w_i$ ;

Return Remind way;

}

## 3.3.2 Modeling of the ProM Agent

The ProM agent performs the following activities: encoding the task and the plan, maintaining the task and the plan, initiating the task, and performing the task, where re

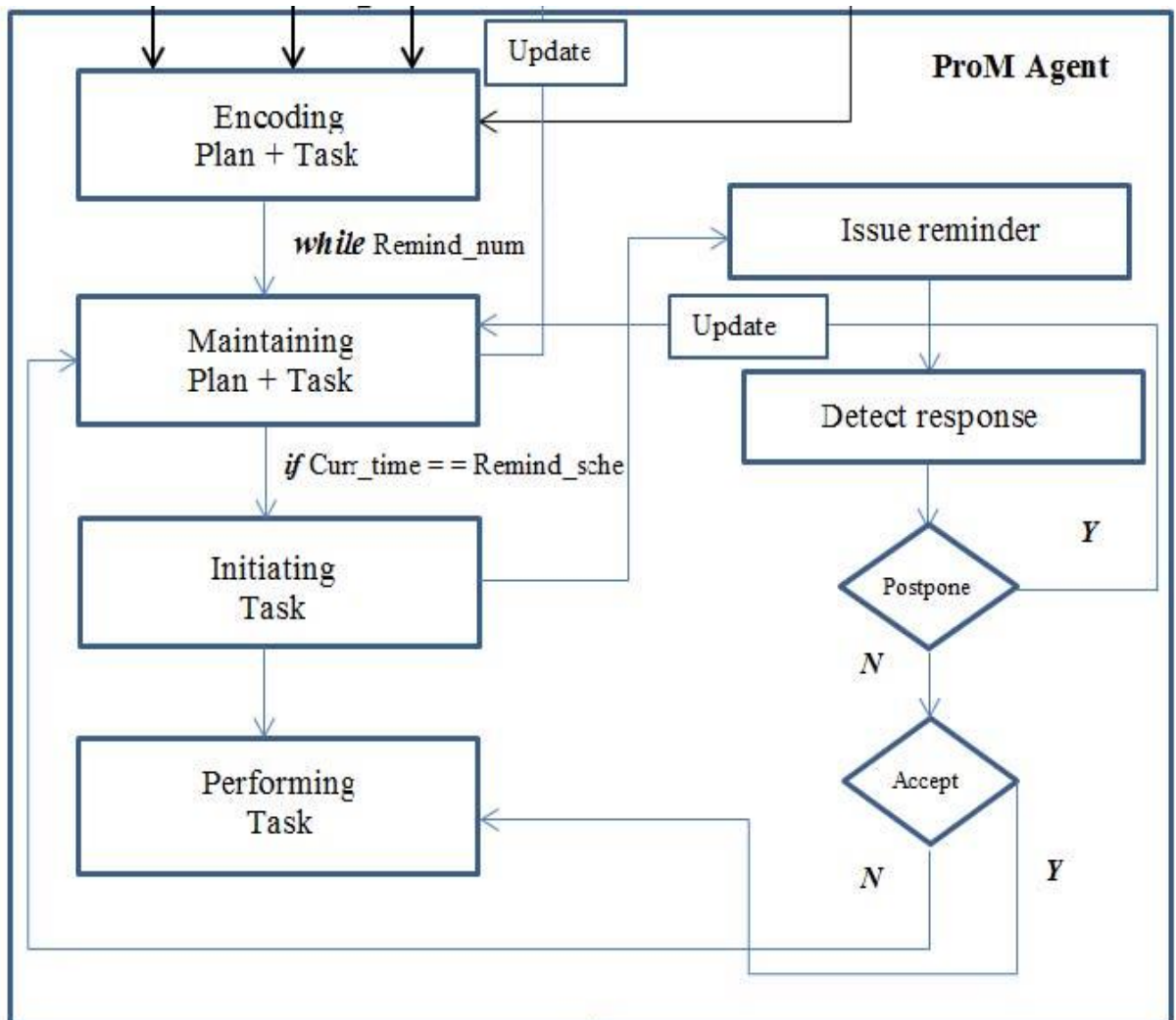
minders are triggered and issued at the initiation stage (see Figure 3.3).

Figure 3.3: The Prospective Memory Agent

(i) Encoding: The encoding consists of five variables, denoted by T ask = En(Remind num, Remind sche, Remind way, W ha, P er), where:

- W ha is the title of the ProM task,
- P er is the person relative to the ProM task,

Remind num, Remind sche, Remind way, W ha, and P er are encoded in the ProM agent.

(ii) Maintaining: Remind num, Remind sche, Remind way, W ha, and P er are maintained in the ProM agent, and waiting for triggering or updating reminders. This stage accepts updates according to users' requirements. It also delivers these updates to the original inputs of the task.

(iii) Initiating:

- Acc and pos are defined as user's responses to the reminder,
- R delay is defined as how long the reminder to be delayed,
- Reminding(Remind way) is defined as the function to issue the reminder accompanying with the way (visual or audio, long or short, music or ring)
- Curr time is defined as the current time;

PSEUDOCODE:

{ Acc = 0; P os = 0; R delay = 0; Reminding(Remind way); i = 0; while ( i < Remind num ) { if (Curr time == $r_i$)

{ Reminding(Remind way);

if ( P os ) then //update the Reminding schedule { update(R delay); goto Retention; } if ( Acc ) then // stop reminders { Remind num = 0;} i++;

 } } }

Please note that the event-based ProM task refers to performing the task by an event. The ProM agent can involve part of the events (e.g., location, person). Two examples illustrate how to initiate event-based ProM tasks: if the user's current location (GPS) is the task location, the system starts to issue reminders, or if the user is calling somebody (in contacts), the system starts to issue reminders.

# Chapter 4 Conclusions and Future Work

## 4.1 Conclusions

We propose to develop a new reminder system to improve the ProM performance. We draw lessons from the existing reminder systems by learning their strengths and overcoming their limitations. At the same time, we analyze the mechanisms of ProM and follow the ProM theory. In this context, our reminder system has been developed with three components: the reminder planer, the ProM agent, and the personalized user model. The reminder planer is responsible for producing the reminder plan according to potential factors on ProM. The ProM agent is responsible for encoding the task and the plan, maintaining the task and the plan, initiating the task, and performing the task, where reminders are triggered and issued at the initiation stage. The personalized user model is responsible for adapting the reminder plan according to human-system interactions.

    To realize the functions of different components of our new reminder model, a series of principles and algorithms are presented in our report. Four potential factors (the complexity of the ongoing task, the importance of the ProM task, the motivation of the ProM task, and user's age) that influence the ProM performance determine the optimal number of reminders. The environmental factors such as the locations of the ProM task and the current task, the objective factor of the user's initial expectation of the reminder's starting time, and the optimal number of reminders determine the optimal reminding schedule. Besides the four factors in determining the optimal number of reminders, another factor (the type of the ProM task, such as personal, work, health, or finance) also determines the

reminding way. The reminders strongly associated with the ProM task type make users retrieve the intention successfully as much as possible. In summary, the three components of the reminder planer, the ProM agent, and the personalized user model constitute our reminder system, which ensures system reliability, and make system optimal and adaptive.

## 4.2 Future Work

We build a reminder model to generate the optimal number of reminders, the optimal reminding schedule, and the optimal reminding way. These results depend on formulating and optimizing a series of factors, such as the user's age, ongoing tasks, the environmental context, and individual differences. Although these factors have been identified to influence the ProM performance and we have known how they influence the ProM performance, we still need to figure out the weight of each factor. Similarly, we also need to figure out the distribution of reminders between the time of starting the reminder and the time of performing the ProM task. These results would help to achieve the optimal reminder plan, which maximizes the probability of remembering to perform the prospective task and minimizes the potential annoyance.

Meanwhile, we propose the user model to learn the user's behaviors and preferences for each feature of reminding, and mediate on the reminder plan to meet the user's preferences as much as possible. We are going to integrate machine learning techniques such as reinforcement learning and supervised learning into study to make our reminder system adaptive in the future.